# Reconciling anomalously fast heating rate in ion tracks with low electron-phonon coupling

Nikita Medvedev[1,2,*], Alexander E. Volkov[3]

[1] Institute of Physics, Czech Academy of Sciences, Na Slovance 2, 182 21 Prague 8, Czech Republic;
[2] Institute of Plasma Physics, Czech Academy of Sciences, Za Slovankou 3, 182 00 Prague 8, Czech Republic;
[3] P.N. Lebedev Physical Institute of the Russian Academy of Sciences, Leninskij pr., 53,119991 Moscow, Russia;
[*] Corresponding author: ORCID: 0000-0003-0491-1090; Email: nikita.medvedev@fzu.cz

## Abstract

Formation of swift heavy ion tracks requires extremely fast energy transfer between excited electrons and a lattice. However, electron-phonon energy exchange is too slow, as known from laser-irradiation experiments and calculations. We resolve this contradiction noticing that electron-phonon coupling is not the sole mechanism of energy exchange between electrons and ions: heating of electrons also alters potential energy surface of atoms, accelerating them and increasing their kinetic energy.

## Introduction

Upon ultrafast high-energy-density deposition into matter, e.g. with femtosecond lasers or swift heavy ions (SHI), an electronic system of a solid is heated to temperatures significantly above the atomic one. This transient state then relaxes leading to equilibration of the electronic and atomic temperatures. This equilibration takes place *via* so-called electron-phonon (or, more generally, electron-ion) coupling. It has been predicted with various theoretical approaches that electron-phonon coupling is a relatively slow process, leading to typically picosecond timescales of equilibration between the electronic and the atomic temperatures at typical conditions of irradiation with SHIs: dynamic structure factor calculations [1,2], a coupled modes approach [3], tight-binding molecular dynamics [4], all show similar results. They are supported by state-of-the-art experimental observations of electron-ion coupling in solids, suggesting that it may be even slower than most theories predict [5,6].

In a stark contrast to those results, it has been known for over decades that SHI irradiation of solids leads to creation of ion tracks at extremely short timescales. SHIs primarily deposit energy into the electronic system, similarly to laser pulse irradiation [7]. The difference from the laser spot is that excited electrons appear within a sub-nanometer vicinity around the ion trajectory [7]. Such an extreme localization results in high gradients of the deposited energy density leading to fast energy transport and cooling of the electronic system in an ion track within hundred-femtosecond timescale – much shorter than that required for electron-phonon coupling to heat the atomic system [1]. Yet, SHI tracks are detected experimentally [8], implying that much faster atomic heating takes place there.

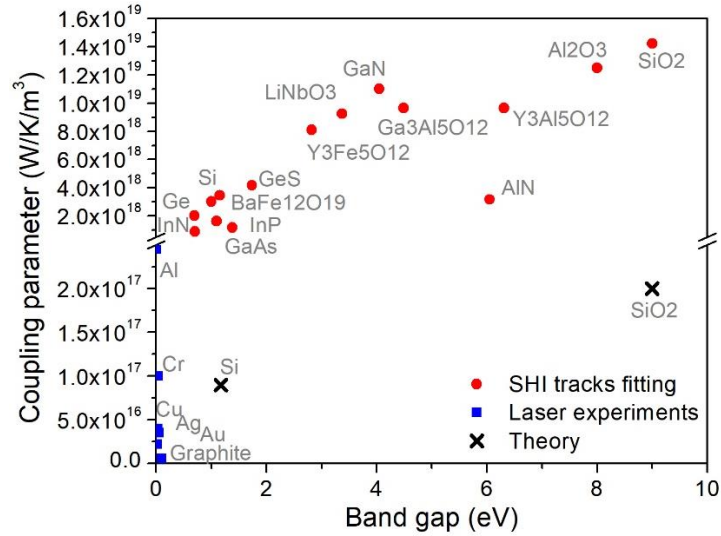

Figure 1. Fitted electron-phonon coupling parameter $G$ in various materials, extracted from the data from Ref. [9], are shown with red circles. Blue squares are electron-phonon coupling parameters measured in laser experiments (taken from Ref. [4] and references therein). Black crosses are calculated coupling parameters at the laser damage threshold doses [4,10].

A necessity of significant heating of materials in SHI impacts to form observable tracks suggests extremely fast coupling of the electron system to the atomic one during times of cooling of the electronic system, see Figure 1 extracted from the data from Ref. [9]. Those estimates imply that the atomic heating must be orders of magnitude faster than that measured in laser-irradiation experiments and consistently calculated with various models. A contradiction ensued, puzzling the respective communities for over two decades[11].

In this work, we point out that the electron-phonon coupling is not the only mechanism of atomic heating. Ultrafast energy deposition into the electronic system also affects an atomic potential energy surface because electrons form an interatomic potential in a solid. Excitation of electrons changes the electronic distribution over the bands (orbitals), which in turn modifies the interatomic potential [12,13]. This converts into increase of the kinetic energy of atoms, because altered interatomic forces push atoms towards their new equilibrium positions [12,14].

At high deposited doses, or correspondingly high electronic temperatures, it may lead to phase transitions even without significant atomic heating. The most famous example of this effect is the so-called nonthermal melting [15,16]. It is known since 1990s to take place in covalently bonded semiconductors [12,14,17]. Recently, it has been theoretically shown that the same effect also takes place in ionic crystals [18,19], oxides [10] and polymers [20]. Thus, nonthermal phase transitions may be regarded as a universal effect taking place in non-metallic targets upon energy deposition faster than the electron-ion (electron-phonon) coupling (while an even more complex picture of various nonthermal effects may take place in metals [21,22]). Such nonthermal phase transitions are accompanied by increase of the atomic kinetic energy (temperature) as a consequence of atomic acceleration as a reaction to interatomic potential changes[20,23].

At doses below a nonthermal phase transition threshold, excited electrons also trigger some "nonthermal" atomic heating *via* modification of the potential energy surface. This effect is known as displacive excitation of coherent phonons [24], or squeezed phonons as a precursor to nonthermal melting [25].

Increase of atomic kinetic energy (atomic heating) due to modification of the interatomic potential after excitation of the electronic system forms a distinct mechanism from the electron-phonon coupling. The latter one relies on the electron transitions between electronic energy levels triggered by atomic displacements, i.e. it is a nonadiabatic effect [4,26,27]. In contrast, nonthermal atomic heating via modification of the interatomic potential is an adiabatic effect, occurring within Born-Oppenheimer approximation [23,25], without nonadiabatic electron-ion coupling involved.

Below we will consider the nonthermal atomic heating in various materials, its rate and timescales at various doses. We will also analyze a synergy between nonthermal heating and electron-ion coupling, considering the dependence of the coupling parameter on the atomic temperature.

## Model

We used XTANT-3 code to model materials response to an ultrafast electronic excitation [28]. It is based on Tight-Binding Molecular Dynamics simulation. Within XTANT-3, electrons populating the valence and the bottom of the conduction band are assumed to adhere to the Fermi-Dirac distribution. This defines the fractional populations of electrons on the energy levels (molecular orbitals) of the material.

Let us point out that after an SHI irradiation, electrons acquire a typical distribution of the so-called 'bump-on-hot-tail' shape (see, e.g., discussion in Ref. [7]): the majority of electrons in the low-energy states are in a close-to-equilibrium conditions, whereas the minority of highly-energetic electrons forms a far-from-equilibrium long tail in the distribution. Although that makes the total electronic distribution highly non-equilibrium, such a bi-modal shape allows us to focus only on the low-energy fraction of electrons. This fraction, located around the SHI trajectory, contains high energy density, and can be described with the Fermi-Dirac distribution, which we will utilize below. The other fraction, the non-equilibrium fast electrons, exhibits ballistic transport outwards from the track core (this topic was studied in detail e.g. in Refs. [7,8,29] and references therein). These electrons will provide atoms with additional kinetic energy due to elastic scattering, as mentioned below, but would not affect the interatomic potential, and are thus excluded from the current consideration.

Transferable tight binding (TB) method describes transient molecular orbitals, interatomic forces depending on the configuration of all the atoms in the simulation box and electronic distribution function, as well as matrix elements for nonadiabatic electron-ion coupling. We apply DFTB method with matsci-0-3 parameterization to describe studied materials [30]. This parameterization uses $sp^3d^5$ LCAO basis set [31]. The TB Hamiltonian and all related properties evolve on each time step of the simulation, thereby tracing evolution of the material parameters as a response to an elevated electronic temperature.

Atomic motion is modeled with Molecular Dynamics (MD) using Martyna *et al.* 4th order algorithm with 0.1 fs time-step [32]. The forces acting on atoms are derived from the potential energy surface provided from the TB method. They also depend on the transient electronic distribution function, thus tracing effects of nonthermal melting due to electronic excitation.

The simulation may be performed either within the Born-Oppenheimer (BO) approximation, or beyond it with nonadiabatic effects included [33]. The Born-Oppenheimer approximation excludes electron-phonon coupling and resulting energy exchange. Non-BO electron-ion coupling is described with the Boltzmann collision integral [4]. The tight binding molecular dynamics calculates matrix elements entering the collision integrals at each time step. Energy exchange from the nonadiabatic coupling of electrons is fed to atoms via velocity scaling[28]. A microcanonical (NVE) ensemble is used with periodic boundary conditions in the simulation box.

The simulated supercells contain 240 atoms for simulation of $Al_2O_3$, and 216 atoms for MgO, $TiO_2$ and $SiO_2$ (quartz). The simulation starts with random velocities assigned to atoms according to the Maxwellian distribution with the room temperature. Atoms are allowed to thermalize for some hundred femtoseconds. Then, the electronic temperature is smoothly increased during 10 fs according to Gaussian function centered at 0 fs.

## Results and discussion

Born-Oppenheimer simulations of electronic and atomic dynamics in $Al_2O_3$, MgO, $SiO_2$ and $TiO_2$ are performed with help of XTANT-3 at various deposited doses.

Figure *2* demonstrates that after the rise of the electronic temperature, the atomic temperature (kinetic energy) starts to increase in all considered materials. The rate of its increase rises with increase of the dose. At doses close to or above the nonthermal damage dose (reported in Ref. [10]), the atomic temperature increase occurs within ~100 fs.

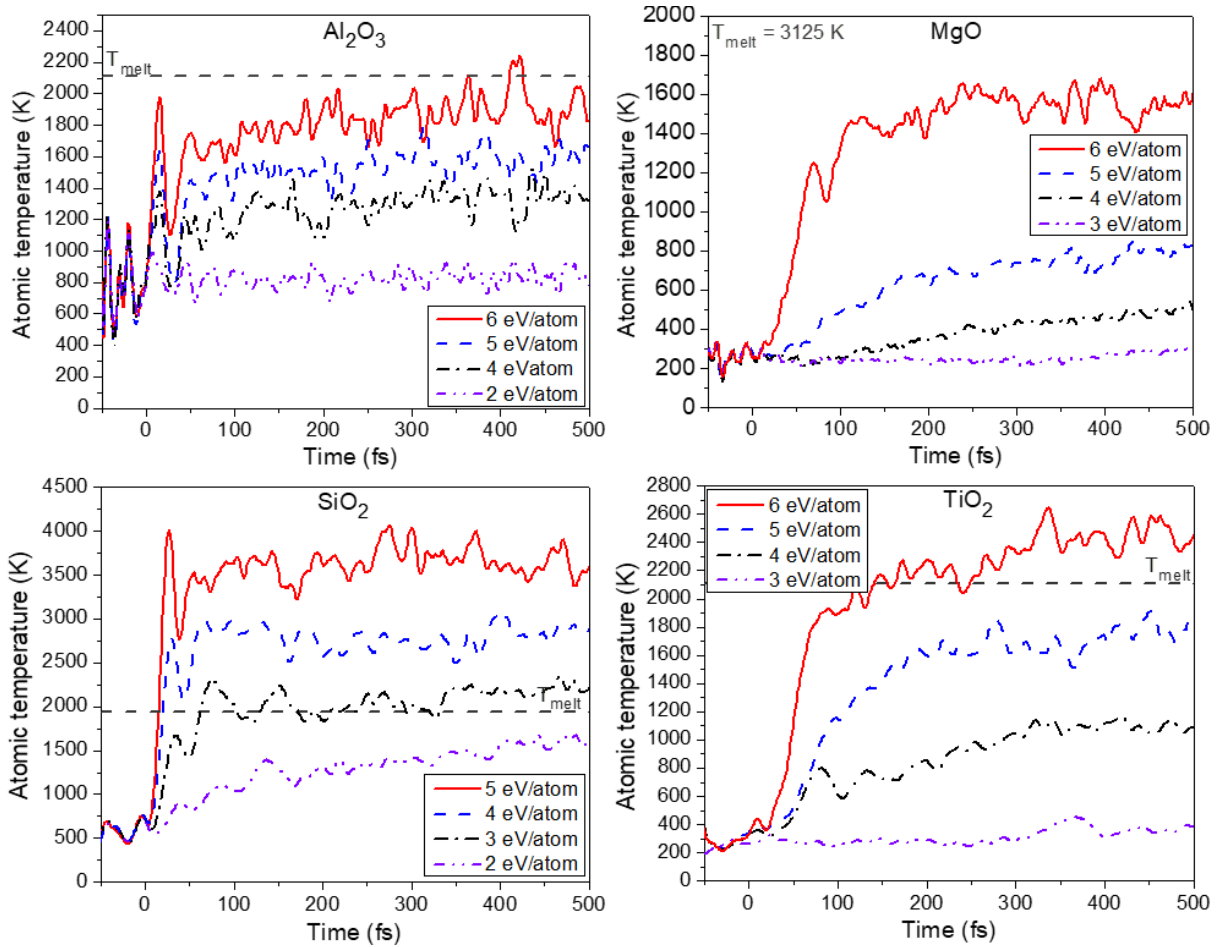

Figure 2. Atomic temperature (kinetic energy) increase after ultrafast energy deposition into the electron system in $Al_2O_3$ (top left panel), MgO (top right panel), $SiO_2$ (bottom left panel), and $TiO_2$ (bottom right panel). Thermodynamic melting temperatures are shown as grey dashed lines for comparison.

We emphasize that this increase of the kinetic energy of atoms is *not* due to electron-ion (electron-phonon) coupling, as this channel of energy exchange is absent within the Born-Oppenheimer simulations. This can be clearly seen in Figure 3a which shows that the electronic and atomic temperatures do not equilibrate within the BO approximation (as was discussed in more detail e.g. in Ref.[34]). Atoms are heated purely due to nonthermal (Born-Oppenheimer) effects – the acceleration caused by changes in the atomic potential energy surface due to electronic excitation. The same effect was reported to take place during nonthermal phase transitions in diamond [23], silicon [34] and polyethylene [20], and experimentally observed in bismuth [35] and silicon [36], suggesting it may be universal in non-metallic targets.

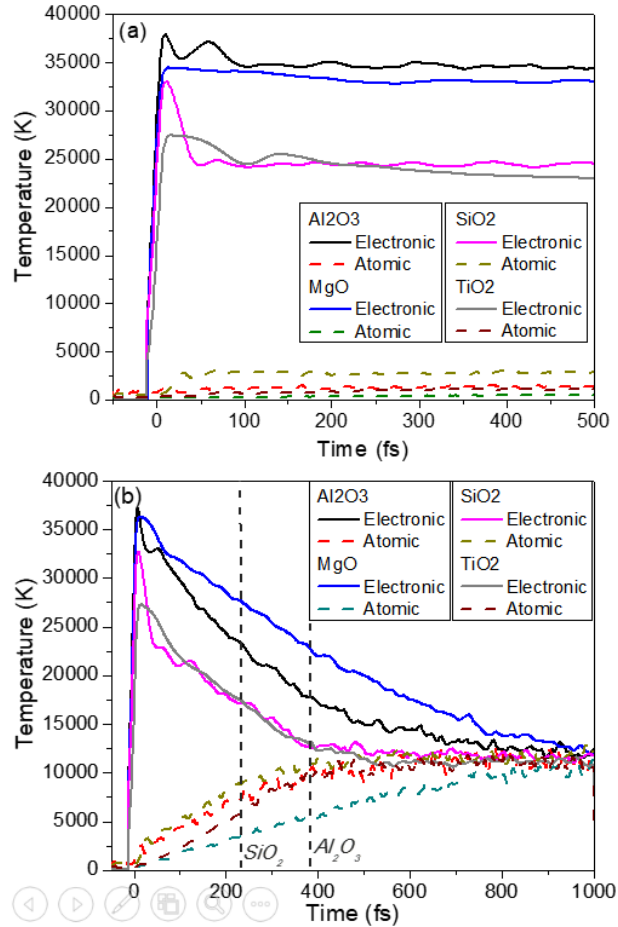

Figure 3. Electronic and atomic temperature changes after ultrafast deposition of 4 eV/atom into the electron system in $Al_2O_3$, MgO, $SiO_2$, and $TiO_2$, modeled (a) within BO (excluding nonadiabatic electron-ion coupling); (b) with electron-ion coupling included. Dashed vertical lines indicate characteristic electron-phonon coupling times extracted from Fig.1.

It is also known that the electron-ion (electron-phonon) coupling parameter depends on both, the electronic and the atomic temperatures (as well as other parameters such as atomic structure and density) [4]. Its dependence on the atomic temperature is nearly linear [4]. A strong kick to atoms due to nonthermal heating thus additionally increases the electron-ion coupling. In turn, this enhanced coupling increases the atomic temperature further, leading to a self-amplifying process.

This synergy results in a very fast increase of the atomic temperature, see Figure 3b. The characteristic times of atomic heating it an SHI track, extracted from Figure 1 (according to the estimate τ=C/G, where C is the atomic heat capacity), are shown for comparison in Figure 3b as vertical lines. We see that our calculated characteristic time of the increase of atomic temperature agrees reasonably well with these mean characteristic times.

In contrast, at lower deposited doses, when electronic temperature is not exceedingly high, atoms do not experience an instantaneous nonthermal kick, and the atomic temperature is not elevated significantly – the coupling is much slower, leading to equilibration over the time of a few (or even up to a few tens of) picoseconds (see e.g. Ref. [10]).

The equilibration of the electronic and the atomic temperatures due to electron-ion coupling that takes place at the picosecond timescale. It will not be relevant to the fast ion tracks problem because of the very fast quenching of electrons due to the fast transport of the electronic energy away from the track core, but is important for laser irradiation of materials [20,34]. In this case of long-lived electronic excitation, at such high doses as presented in Figure 3, much of the deposited energy is spent on electronic heating and changes in the electronic structure causing changes in the potential energy of the atomic system as well as overcoming the barriers of atomic disorder. The remaining part of the deposited energy converts into the kinetic energy of atoms – the increase of their temperature up to ~10 kK. At such temperatures, the system forms the so-called warm dense matter (WDM) state [37–39]. Creation of WDM has been observed experimentally in ultrafast laser irradiation experiments [40,41].

The results demonstrated that the atomic heating at the doses above a few eV per atom can take place within a hundred femtoseconds. Energy densities delivered by fast ions may transiently reach even higher doses [8,29]. The timescales of this atomic heating are comparable with the lifetime of the excited electrons and valence holes in SHI tracks. Monte Carlo simulations showed that electrons cool down and travel outwards from the track core within ~ 100 fs, but the valence holes, which are typically heavier than conduction band electrons, last longer in the track core [42]. Those times may be sufficient if not for a completion of a nonthermal phase transition, at least for significant nonthermal energy transfer from the electronic to the atomic system. Adding to it the energy transferred to atoms due to scattering of fast ballistic (nonthermalized) electrons, having energies up to tens of keV, can induce structure transformations and form tracks in many materials even without slow electron-phonon coupling [29,43].

It thus resolves the discrepancy between the anomalously fast heating of atoms in SHI tracks and much too slow electron-phonon coupling: heating of atoms in a track takes place mainly *via* combined channels of nonthermal atomic heating and scattering of ballistic electrons, not *via* electron-phonon coupling.

Thus, estimates of the "electron-phonon coupling" extracted from the fast ion tracks parameters, such as in Figure 1, must be interpreted as reflecting the rate of nonthermal atomic heating (increase of kinetic energy) and not as a real electron-ion (electron-phonon) coupling parameter.

A similar notion of heating of atoms via mechanisms other than electron-phonon coupling has been previously discussed in terms of the Coulomb explosion model [44]. The major shortcoming of that suggestion was that the Coulomb explosion is known to take place in small molecules and nanoclusters, or possibly thin layers or surface of materials, but is dubious in solids [7]. Valence holes in solids may transport out of the track, and a massive amount of electrons in the surrounding solid neutralizes the uncompensated charge around an ion trajectory within a few femtoseconds [45]. In contrast, nonthermal effects do not require an unbalanced charge. Born-Oppenheimer heating of atoms is thus a universal effect in nonmetallic materials, which must be accounted for in models of any type of ultrafast electronic excitation.

# Conclusions

In conclusion, the estimates of the electron-phonon coupling parameter extracted from the data on swift heavy ion track sizes do not reflect the electron-phonon coupling *per se*, but must be interpreted as a quantity reflecting the nonthermal heating of atoms *via* modifications of the potential energy surface due to electronic excitation. This notion reconciles the much debated problem of extremely fast atomic heating in swift ion impacts (on hundred femtosecond timescales) with slow electron-phonon coupling (acting on picosecond timescales).

Our results suggest that the nonthermal heating of atoms is a universal effect, which presents in nonmetallic crystalline materials under ultrafast energy deposition, and thus must be taken into account in appropriate models. The effect is especially pronounced at doses above the nonthermal damage threshold. It leads to extremely fast increase of the atomic temperatures, on the order of a hundred femtoseconds. This effect is not limited to ion impacts but is important for all kinds of ultrafast energy deposition into the electronic system and is expected to play a role, e.g., in femtosecond laser ablation.

## Conflict of Interest

The authors declare no conflict of interests, financial or otherwise.

## Data Availability Statement

The data that support the findings of this study are available from the corresponding author upon reasonable request.

## Acknowledgements

Computational resources were supplied by the project "e-Infrastruktura CZ" (e-INFRA LM2018140) provided within the program Projects of Large Research, Development and Innovations Infrastructures. NM gratefully acknowledges financial support from the Czech Ministry of Education, Youth and Sports (Grants No. LTT17015 and LM2018114). This work benefited from networking activities carried out within the EU funded COST Action CA17126 (TUMIEE) and represents a contribution to it. Partial support from project No. 16 APPA (GSI) funded by the Ministry of Science and Higher Education of the Russian Federation, is acknowledged by AV.